\documentclass[11pt,letterpaper,twocolumn]{article}

\usepackage{times}
\usepackage[utf8x]{inputenc}
\usepackage[T1]{fontenc}
\usepackage{listings}
\usepackage{float}
\restylefloat{table}

\usepackage[letterpaper,top=3cm,bottom=3cm,left=2cm,right=2cm,marginparwidth=1.75cm]{geometry}
\usepackage{amsmath}
\usepackage{graphicx}
\usepackage[colorinlistoftodos]{todonotes}
\usepackage[hyphens]{url}
\usepackage{algorithm}
\usepackage{algorithmic}
\usepackage{indentfirst}

\title{\huge{\textit{Dancing with Donald}}\\ \Large Polarity in the 2016 Presidential Election\footnote{https://github.com/darrenmei/election\_mechanisms}}
\author{Robert Chuchro, Darren Mei, Kyle D'Souza\\
{\tt\small chuchro3@stanford.edu, dmei@stanford.edu, kvdsouza@stanford.edu}}

\date{December 7, 2018}

\begin{document}
\maketitle

\begin{abstract}
\textit{In almost every election cycle, the validity of the United States Electoral College is brought into question. The 2016 Presidential Election again brought up the issue of a candidate winning the popular vote but not winning the Electoral College, with Hillary Clinton receiving close to three million more votes than Donald Trump. However, did the popular vote actually determine the most liked candidate in the election? In this paper, we demonstrate that different voting policies can alter which candidate is elected. Additionally, we explore the trade-offs between each of these mechanisms. Finally, we introduce two novel mechanisms with the intent of electing the least polarizing candidate.}

\end{abstract}
\section{Introduction}
In 1970, Salvador Allende, head of the Popular Union party in Chile, became the first Marxist ever elected to the national presidency of a democracy. The only problem: the election mechanism used the plurality rule. Allende won without a majority, having only 36.2\% of the vote and around 60\% against. \cite{chile} The following years were marked by significant protests. Supported by the U.S. Government, a military coup followed. The result was the fall of one of the world's strongest democracies to a 17-year military coup (1973-1990). While the non-majority pluralistic vote outcome was not the sole cause of the 17-year dismantlement of democracy, the election mechanism design was certainly a strong factor in creating the public sentiment that supported and allowed a military coup to take root.\\
\noindent
\textit{Why do election mechanisms matter?}

Election mechanisms are of great importance to governments. Elections dictate the future of nations, and democratic countries often have very different mechanisms from each other. The rise of the Arab Spring and increased criticism on polarization in America, potentially due to our two-party election system, has generated renewed focus on election mechanisms. \cite{Norris} Additionally, while there have been recent movements (i.e. the recent new primary system in California as of 2012) to reduce polarization and moderate recent elections, preliminary studies show an inconclusive change in resulting outcome. \cite{california}

Electoral systems are one of the crucial elements of a well-functioning democracy. \cite{rickard} Moreover, because electoral systems influence important political outcomes, they impact legislative behavior, electoral competition, the relationship between votes and seats in parliaments, and the number of parties in government. \cite{rickard} For example, our two-party system is a direct result of our plurality system. \cite{levin} Changing our election mechanism could result in a completely different party dynamic and new representatives, not to mention a complete change of the millions of dollars of infrastructure built under the two-party system. Finally, election mechanisms can potentially impact the direction of policies. By electing candidates of different polarity or incentivizing candidates to pander towards certain populations or demographics, these mechanisms can potentially influence policy and thereby impact the lives of millions. As shown by Chile's 1973 presidential election, a country's election mechanism can either bolster or endanger the very fabric of its democracy.

\section{Problem Statement}

Motivated by the theory that different election systems can affect the outcome of elections, we investigate how election systems can specifically affect the election of polarized vs. moderate candidates. Since there was no existing ranked list of American citizens preferences from the 2016 election results, we collected our own dataset, surveying over 750 U.S. citizens. Subsequently, we normalized our data and put it to the test, running several election mechanisms to reach preliminary conclusions on which systems favor certain candidates over others.

As explained in the following section, polarization and radicalization is an increasing topic of interest in Political Science. However, we chose this project because we saw a gap between a very interesting topic and Computer Science's emphasis on mechanism design. By collecting our own dataset and applying mechanisms learned in this class and elsewhere, we hope this can be a significant first step in analyzing the effect of voting mechanisms on election results, and ultimately, their policy implications. Following this, we plan to share these results and further collaborate with professors in the field. 

\section{Related Work}

We approached our literature review from two angles, 1) voting mechanism design and 2) research on polarization: \\\\
\textit{Voting Mechanisms}

Existing research shows that given the same race, different mechanisms can elect different candidates. \cite{lecture3} Kenneth Arrow, Nobel Prize winning Economist, began work on this with the RAND Corporation in 1951, examining social choice theory. Arrow's paradox showed that when voters have three or more distinct options, there exists no ranked voting electoral system that can both convert the ranked preferences of individuals into a community-wide ranking while satisfying the following conditions: unrestricted domain, non-dictatorship, independence of irrelevant alternatives and Pareto efficiency. \cite{arrow} Voting mechanisms have been studied to various degrees, with mechanisms offered by Marquis de Condorcet \cite{condorcet}, A.H. Copeland \cite{copeland}, Clyde Coombs \cite{coombs}, and Duncan Black \cite{black}. Moreover, research has been done on more advanced / complicated mechanisms like minimum violations \cite{minviolations}, ranked pairs \cite{myerson}, min-max rule, and the Kendall-Wei power rankings, which are often used in sports. Nobel Laureates, computer scientists, and economists have conducted significant research on voting mechanisms. Nevertheless, as of today, work remains with regard to an analysis on election mechanisms' affect on reducing polarization.\\\\ 
\textit{Polarization}

Similarly, in political science, many scholars have researched polarization and its increase in American politics. However, less focus has been placed on how different voting mechanisms would result in different voting outcomes.

Significant research has been done by Stanford University scholars like Andy Hall, Shanto Iyengar, Bruce Cain, and David Brady, on polarization and election systems. Hall, Political Science Professor at Stanford, claims that "U.S. legislatures have become increasingly polarized and dysfunctional in part because of how difficult running for and holding office is." Combining theoretical and empirical evidence, he "shows how voters are forced to elect extremists because moderates don't run for office, and document[s] how the rising costs of running for office, and the falling benefits of holding office, have deterred moderates from running."\cite{hall} Moreover, Hall has researched results in primaries \cite{hallextremeprimaries}, and concludes that extremists and their affiliated party tend to suffer penalties in the general election. To Hall, the general election system tends to have a moderating effect on elections in the House of Representatives in America. Similarly, Shanto Iyengar, another Political Science Professor, provides compelling evidence to support growing polarization in America. \cite{iyengar} Iyengar's research raises important questions, many of which are beyond the scope of this paper. 

Finally, there exists significant academic consensus on the importance of election systems on policy outcomes. Amartya Sen's work on social choice \cite{sen} sets up a compelling argument on the benefits and drawbacks of majority voting in deciding on issues of poverty and welfare. Additionally, Rickard puts significant effort in analyzing the impact of electoral systems on policy outcomes, but fails to reach significant conclusions on how or why this impact exists. \cite{rickard} With this significant backdrop and consensus that election systems help define policy outcomes, we turn our focus to analyzing the impact of election mechanisms on electing moderate or polarized candidates. 

\section{Methods}
\subsection{Data Collection}

Since the U.S. Presidential election does not have voters provide a ranked list of candidate preferences, we decided to collect our own data set for the purpose of this project. The reasons were twofold: we did not see similar compelling data on an individual scale, and we believed this data set would be a helpful first step for our research as well as future publications by others. While we could have chosen (and plan on surveying other elections) we selected the USA election because of the simplicity and reliability of garnering high user numbers and building a first proof of concept. 

In terms of specific methodology, we set up a Mechanical Turk account and surveyed over 700 respondents, all of which were domiciled in America and represented a fairly diverse set of geographical respondents. As one of our members has used mTurk in prior research investigating a causal link between the anti-vaccination movement and Donald Trump's media presence, we had good experience in designing HIT tests. We believe that we received genuine responses; however, there are significant risks in assumption. For instance, users could have evaded mTurk's Bot Detection systems or could have answered dishonestly or randomly. Nevertheless, for a first crack, our paper focuses more on the mechanism design and leaves validation of results to subsequent, better funded research projects.

Our survey, entitled US 2016 Presidential Election Preferences, took on average 2 minutes and 11 seconds, asked 8 questions, and was powered by Qualtrics. Our first five questions surveyed demographics (age, gender, race, party preference, and whether they voted in 2016). Subsequently, we ask participants to rank 8 candidates (Gary Johnson, Jill Stein, Donald Trump, Marco Rubio, Bernie Sanders, Ted Cruz, Gary Johnson, John Kasich), and to specify the last candidate rank they thought would make an acceptable President. We used this question for measuring approval voting \cite{levin} and our newly defined "Polarization Rule". Finally, we ask voters to select a candidate for the first round of a plurality with runoff election (French election) setup. This question yielded interesting results, which we address later on.\newline
\newline
\textit{Ethical Considerations}

As an important aside, we encountered two major ethical considerations regarding our usage of Mechanical Turk and Qualtrics. We spent 3 cents per survey response (2 cents to Turker, 1 cent to mTurk), due in part to making our survey keywords attractive (easy, quick, survey, demographics, Stanford, voting). It was remarkable to get potent data for such a reasonable price, and we got universally positive feedback for keeping our survey short and contained. However, the fact remains that these surveyors were paid an average of \$0.60 an hour, which does not come close to a proper working wage.

Secondly, our Qualtrics survey also released confidential information, specifically the exact latitude and longitude of the surveyed location. When we plugged in our survey geographics into Google Maps to ascertain how diverse our surveyed population, we had the ability to see with Street View the actual house in which the survey was completed. In the future, researchers, Qualtrics, and Mechanical Turk should be more careful to address the potential privacy risks of completing similar surveys to ours.

\subsection{Accounting for Voter Demographics}
\indent Though we collected a large number of survey responses through Mechanical Turk, the population that fills out Mechanical Turk surveys does not represent the general US population. In general, we found that the average Mechanical Turk user skews younger and more likely to be a Democrat or Independent than the average US voter. To account for this difference, we included a demographic information section in our survey to identify voters. In addition, we asked participants to also indicate whether they had voted or not in the 2016 election. Using these survey answers, we were able to determine the demographic breakdown of our survey participants and normalize result. \\
\indent
When writing the survey, we wanted our questions to align with the demographics outlined in the 2016 Election Exit Polls.\cite{exitpoll} In doing so, it was easy to determine how our participants compared to the voters in the 2016 election. After determining the demographic breakdown, we then calculated a correction constant based on the representation of a demographic in the 2016 election divided by the representation of the same demographic in our survey. We then combined the correction constants across all of the demographics we collected to weight each of our participants' votes correctly. With this, we hoped to make our results more representative of the 2016 voter population and correct for potential participant bias. However, one demographic we did not sample for was participant income. The impact of this decision will be further discussed in later sections.

\subsection{Proposed Mechanisms}
\subsubsection{Existing Mechanisms}
\noindent
\textit{Simple Plurality}\\
\indent
The most basic voting mechanism is when each voter submits a single vote for a candidate and the winner is the candidate with the most votes. Most of the elections in the United States utilize this simple plurality voting rule, and it is one of the reasons for the emergence of the two-party system.\cite{levin} This mechanism is not strategy-proof, as voters have more incentive to vote for the candidate they like the most that also has a high chance of winning instead of the candidate that they might actually prefer. \\

\noindent
\textit{Plurality with Runoff}\\
\indent
An extension of the simple plurality mechanism is to have an initial round similar to a simple plurality vote where voters submit a single vote. However, a winner is not decided unless a candidate holds the majority of the votes. If there is no clear winner after this initial round of voting, a runoff is held between the two candidates with the highest number of votes. In this second round, voters then choose either of the two potential candidates. Although the second round does determine the truthful preferred candidate for voters, voters have an incentive in the initial round to vote for a candidate they believe will end up in the final runoff. \\

\noindent
\textit{Ranked Choice}\\
\indent
In a ranked choice voting mechanism, also known as single transferable vote or instant-runoff voting, voters submit a ranked list of the candidates instead of just their first choice. If a candidate receives the majority of the first place rankings then they win the election. Otherwise, the candidate with the least amount of first-place rankings is removed from the election and the lists are reconsidered excluding them. This process continues until one candidate receives a majority of the votes, and they are declared the winner. This mechanism is also not strategy-proof, but it is tougher to game ranked choice voting than it is to game the two plurality methods discussed above. Some ranked choice mechanisms don't require voters to rank all of the candidates in the election, but for this project, we assume that all of the candidates are ranked.\cite{lecture3} \\

\noindent
\textit{Coombs' Method} \\
\indent
A similar method to Ranked Choice or Instant-Runoff Voting is Coombs' Method.\cite{coombs} Voters still submit a ranked list of their choices, but instead of eliminating candidates based on who has the least number of first-place votes, Coombs' Method instead eliminates candidates based on who has the most last-place votes. After removing the candidate with the most last-place votes, the lists are recomputed similarly to Ranked Choice voting. This process continues until there is one candidate remaining in the election, who is declared the winner. Though the process seems similar to Ranked Choice voting, eliminating the candidate that is least liked (i.e. has the most number of last place votes) can result in a much different outcome. 

\noindent
\textit{Borda Count}\\
\indent
The Borda count voting mechanism also has voters submit their ranked choices, but instead of removing candidates like the ranked choice mechanism it computes a score. For each ranked list a candidate receives a score based on how they were placed in the list. If a candidate is the first choice, then they receive a score of $|A|$ where $|A|$ represents the number of candidates in the election. If the candidate is the second choice, they receive a score of $|A| - 1$. This continues until the last choice candidate, who receives a score of $1$. The scores are then summed up across all of the voters' lists, and the candidate with the highest Borda count wins the election. Again, this mechanism is also not strategy-proof, as voters are incentivized to rank the closest competitor to their preferred candidate last instead of at their truthful ranking.\cite{lecture3}

\noindent
\textit{Approval Voting}\\
\indent
This mechanism asks voters to submit a set of candidates that they approve of instead of ranking all of the candidates. If candidates are within this set, then they receive a vote. The votes are then summed across all of the submitted sets of approved candidates, and the candidate with the most votes is deemed the "most approved" and wins the election.\cite{pacuit} Again, this is not strategy-proof because voters will not include competitors to their top choice in their approval set even if they truthfully approve of them. \\

\noindent
\textit{Copeland Rule} \\
\indent
The Copeland Rule deals with pair-wise victories and is used to determine the Condorcet winner.\cite{copeland} A formal definition of a Condorcet winner and its presence in various voting rules is discussed in Section 5. In this voting rule, each voter submits their ranked choices. However, instead of tallying a score or removing candidates based on some policy, the candidates are ordered in terms of their head to head victories against the other candidates minus their head to head defeats. A head to head victory is found if, throughout all of the ranked lists, a candidate is ranked higher than another candidate a majority of the time. 

\subsubsection{Novel Mechanisms}
In addition to implementing the existing mechanisms described above, we also designed two new mechanisms with the aim of achieving a different election outcome. \\

\noindent
\textit{Moderation Rule}\\
\indent
In this rule, we combine standard Ranked Choice voting with Coombs' rule, alternating which candidate to return. Ideally, this would balance out removing candidates that are not the most liked and also the most disliked. The rule is as follows.

\begin{algorithm}[h]
\caption{Moderation Rule}
\begin{algorithmic}
\WHILE{more than one candidate remains}
\STATE Calculate first-place vote totals
\STATE Remove candidate with least first-place votes
\IF{one candidate remains}
\RETURN remaining candidate
\ENDIF
\STATE Recompute ranked lists
\STATE Calculate last-place vote totals
\STATE Remove candidate with most last-place votes
\STATE Redo ranked lists without the removed candidate
\ENDWHILE
\RETURN remaining candidate
\end{algorithmic}
\end{algorithm}

Though this rule decides to start by removing the candidate with the least first-place votes, a variation could start by removing the candidate with the most last-place votes instead. \\

\noindent
\textit{Polarization Rule}\\
\indent
We created the Polarization Rule with the intent of electing the candidate for whom voters would have the least negative reaction. To do this, we made the policy outlined in Algorithm 2.

\begin{algorithm}[h]
\caption{Polarization Rule}
\begin{algorithmic}
\FOR{every voter ranked list}
\STATE $i \xleftarrow[]{}$ index of last acceptable candidate in list
\FOR{index $j$ where $i < j \leq$ number of candidates}
\STATE{candidate score -= $ (j - i)^2$}
\ENDFOR
\ENDFOR
\RETURN candidate scores
\end{algorithmic}
\end{algorithm}

This policy decreases a candidates score the further away they are from a voter's acceptable threshold. The decision to square this distance was made because we wanted to emphasize the increase of being the last place in a voter's ranked list. The candidate with the least negative final score will be declared the winner, because they will have accumulated the lowest negative reaction from voters. Ideally, the result of this type of election would be met with the least resistance because the average voter would be the least upset about the elected candidate winning.

\section{Properties of Mechanisms}
\noindent
\textit{Condorcet Condition}\\
\indent
Before discussing the properties of each mechanism, the concept of a Condorcet winner should be brought up. Theorized by Marquis de Condorcet in 1785, the Condorcet winner is a candidate that defeats all other candidates in head-to-head victories.\cite{condorcet} These head-to-head victories are determined by summing up the number of times a candidate is ranked higher than another candidate in voters' choices and seeing which candidate is ranked higher the majority of the time. The Condorcet winner can be seen as the most liked candidate in the election, because they are the candidate that is on average more preferred than every other candidate. However, there does not necessarily have to be a Condorcet winner in every election, as shown in Table 1.
\begin{table}[h!]
\centering
{\begin{tabular}{|c | c | c|} 
 \hline
 Voter 1 & Voter 2 & Voter 3 \\ [0.5ex] 
 \hline
 A & B & C\\ 
B & C & A \\
C & A & B\\
 \hline
\end{tabular}}
\caption{Election without Condorcet Winner}
\label{table:time_per}
\end{table}
\\
\indent
In this example, there are three candidates, A, B, and C, and three voters. As seen in their ranked lists above, Candidate A wins the head-to-head matchup against B but loses to C, and B wins the head-to-head matchup against C. This means that there is no clear Condorcet winner that wins against all other candidates. \\
\indent
Since a Condorcet winner is not guaranteed, Condorcet elections require some tiebreaker method in the case where there is a tie. However, when a Condorcet winner does exist, it could be argued that any mechanism should select the Condorcet winner.\cite{levin} We will see, however, that many of the mechanisms reviewed in this project do not satisfy the Condorcet condition, meaning that they don't necessarily choose the Condorcet winner if one exists. \\

\noindent
\textit{Existing Mechanisms and the Condorcet Condition}\\
\indent
Previous work has shown that the plurality rule, ranked-choice voting, and the Borda count do not satisfy the Condorcet Condition, as there are examples of elections where a Condorcet winner does not win the election.\cite{homework} Since the plurality rule does not satisfy the Condorcet Condition, it extends from that that the plurality with runoff also does not satisfy the Condorcet Condition. The same argument can be said for ranked-choice voting not satisfying the Condorcet Condition and Coombs' Method not satisfying the Condorcet Condition, because Coombs' Method is just a variation on how to remove the candidates. \\
\indent
The approval voting method also does not satisfy the Condorcet Condition, which can be seen in Table 2. The underlined candidate represents the last acceptable candidate in a voter's ranked list.
\begin{table}[h!]
\centering
{\begin{tabular}{|c | c | c | c | c|} 
 \hline
 Voter 1 & Voter 2 & Voter 3 & Voter 4 & Voter 5\\ [0.5ex] 
 \hline
 \underline{A} & \underline{A} & \underline{C} & \underline{C} & B\\ 
B & B & B & B & \underline{A} \\
C & C & A & A & C\\
 \hline
\end{tabular}}
\caption{Approval Voting Example}
\label{table:time_per}
\end{table}
In this example, B is the Condorcet winner, but A wins the election because it is included in the most approval voting sets. \\

\begin{figure*}[h!]
    \centering
    \includegraphics[scale=0.4]{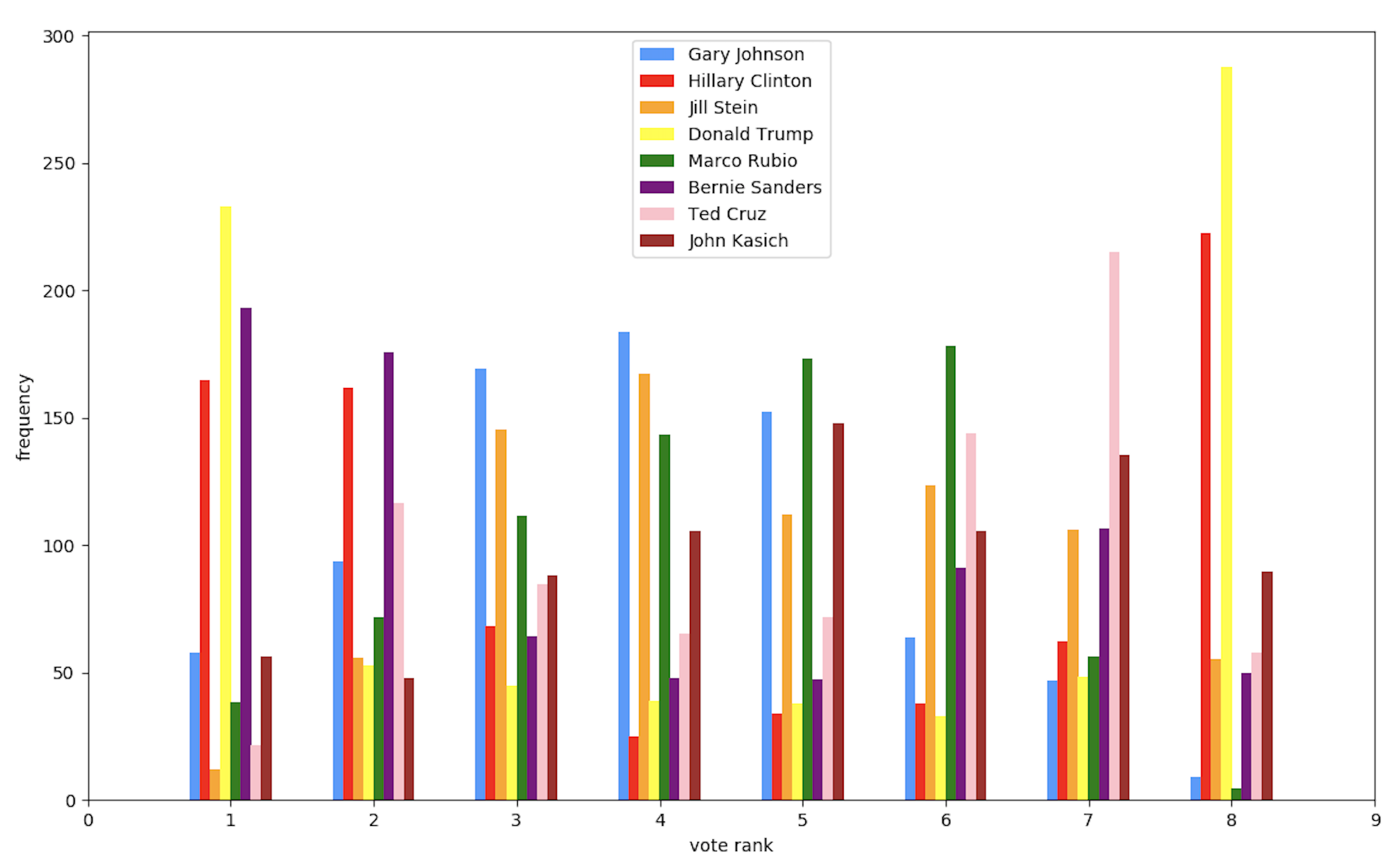}
    \caption{Aggregate of ranking per candidate}
    \label{figure:voter_rankings}
\end{figure*}

\indent
The Copeland Rule satisfies the Condorcet condition. To prove this, assume there is a Condorcet winner in the election. This means they will have won $|A| - 1$ matchups against all of the other candidates. Any other candidate could have won at most $|A| - 2$ matchups because they would have lost to the Condorcet winner. Since the Copeland Rule computes a winner based on candidates' head-to-head matchups, then the Condorcet winner will win the election because they are guaranteed to have more head-to-head victories. Therefore, the Copeland Rule satisfies the Condorcet condition.\cite{cmulecture}\\

\noindent
\textit{Novel Mechanisms and the Condorcet Condition}\\
\indent
The Moderation Rule that was proposed in the previous section does not satisfy the Condorcet Condition. See the following election described in Table 3 for a counterexample.
\begin{table}[h!]
\centering
{\begin{tabular}{|c | c | c | c | c|} 
 \hline
 Voter 1 & Voter 2 & Voter 3 & Voter 4 & Voter 5\\ [0.5ex] 
 \hline
 A & A & C & C & B\\ 
B & B & B & B & A \\
C & C & A & A & C\\
 \hline
\end{tabular}}
\caption{Moderation Rule Example}
\label{table:time_per}
\end{table}
\\
\indent This example resembles the Approval Voting Example in Table 2, but there are no acceptable boundaries. Candidate B is still the Condorcet Winner, but they are removed in the first round of the Moderation Rule because they have the least number of first-place votes. 

\indent
The Polarization Rule also does not satisfy the Condorcet condition. A counter-example shows this in Table 4. Similar to the Approval Voting Example, the last acceptable candidate for each list is underlined. \\
\begin{table}[h!]
\centering
{\begin{tabular}{|c | c | c |} 
 \hline
 Voter 1 & Voter 2 & Voter 3\\ [0.5ex] 
 \hline
 B & B & A \\
 A & A & \underline{C} \\
 \underline{C} & \underline{C} & B\\ 
 \hline
\end{tabular}}
\caption{Polarization Rule Example}
\label{table:time_per}
\end{table}

\indent In this example, B is again the Condorcet winner, but since they are the only candidate that is ever outside the acceptable boundary, they receive the highest polarization score and are therefore guaranteed to lose.

\section{Results}


\subsection{Overview}
Based on every mechanism we tested, the majority result is that Bernie Sanders is the winner of the 2016 election (Table \ref{table:all_rankings}). Based on our data, Bernie Sanders is also a Condorcet Winner, since he wins every single matchup in Copeland's Rule.(Table \ref{table:copeland}) \\
\indent One of the elections which Bernie Sanders loses is in our proposed Polarization Rule, which penalizes Bernie Sanders for having a significantly higher number of low ranks compared to candidates Gary Johnson and Marco Rubio (Table \ref{figure:voter_rankings}).

\subsection{Plurality Results}
The only other case where Bernie Sanders loses is in the Simple Plurality mechanism (Table \ref{table:sp}), where the winner is Donald Trump. While this is the simplest mechanism to understand and implement, the trade-off is that the mechanism does not take into account other factors, such as the overwhelming amount of voters who ranked Donald Trump last. (Table \ref{figure:voter_rankings})
    \begin{table}[h!]
    \centering
    {\begin{tabular}{| c || c | c |}
           \hline
     Ranking & Candidate & Score \\
       \hline
1 & Donald Trump & 232\\
2 & Bernie Sanders & 192\\
3 & Hillary Clinton & 162\\
4 & Gary Johnson & 57\\
5 & John Kasich & 56\\
6 & Marco Rubio & 38\\
7 & Ted Cruz & 21\\
8 & Jill Stein & 12\\\hline
    \end{tabular}}
    \caption{Simple Plurality}
       \label{table:sp}
       \end{table}

\subsection{Plurality with Runoff Results}
    The top two candidates selected after the initial round are Bernie Sanders and Donald Trump, followed by a close showdown which results in a Bernie Sanders victory by 3 votes.
\\ 
\indent It is worth noting that the initial round votes submitted for this question on the survey are different than the top ranks submitted by voters (see table \ref{table:sp} vs table \ref{table:pwr}). The issues of truthfulness and strategic voting will be addressed in the subsequent section.

    \begin{table}[h!]
    \centering
    {\begin{tabular}{| c || c | c | c |}
           \hline
     Ranking & Candidate & Score & Runoff Score\\
       \hline
1 & Bernie Sanders & 224 & 336\\
2 & Donald Trump & 201 & 333\\
\hline
3 & Hillary Clinton & 164 & N/A\\
4 & John Kasich & 69 & N/A\\
5 & Marco Rubio & 44 & N/A\\
6 & Gary Johnson & 39 & N/A\\
7 & Ted Cruz & 21 & N/A\\
8 & Jill Stein & 8 & N/A\\\hline
    \end{tabular}}
    \caption{Plurality with Runoff}
       \label{table:pwr}
       \end{table}

\subsection{Approval Voting Results}

Bernie Sanders is included as an acceptable candidate by most voters, and he therefore wins the Approval Voting election. Out of all mechanisms, Approval Voting is a fairly close contest between all 8 candidates.

    \begin{table}[h!]
    \centering
    {\begin{tabular}{| c || c | c |}
           \hline
     Ranking & Candidate & Score \\
       \hline
1 & Bernie Sanders & 265\\
2 & Hillary Clinton & 249\\
3 & Donald Trump & 230\\
4 & Gary Johnson & 212\\
5 & Marco Rubio & 198\\
6 & Ted Cruz & 190\\
7 & John Kasich & 171\\
8 & Jill Stein & 157\\\hline
    \end{tabular}}
    \caption{Approval Voting}
       \label{table:av}
       \end{table}




\clearpage

\subsection{Ranked Choice Results}

While Ranked Choice does not penalize low ranks as directly as Coombs' Method, in further iterations of the algorithm, low ranks will remain low, while middle ranks are more likely to be pushed towards the top.

    \begin{table}[h!]
    \centering
    {\begin{tabular}{| c || c | c |}
           \hline
     Ranking & Candidate & Score \\
       \hline
1 & Bernie Sanders & 8\\
2 & Donald Trump & 7\\
3 & Hillary Clinton & 6\\
4 & Gary Johnson & 5\\
5 & John Kasich & 4\\
6 & Marco Rubio & 3\\
7 & Ted Cruz & 2\\
8 & Jill Stein & 1\\\hline
    \end{tabular}}
    \caption{Ranked Choice}
       \label{table:rc}
       \end{table}

\subsection{Coombs' Method Results}

Coombs' Method penalized candidates with many low ranks, which is why we see Hillary Clinton and Donald Trump at the bottom of the list. The intuition behind this mechanism is that it will elected the "least hated candidate." 

    \begin{table}[h!]
    \centering
    {\begin{tabular}{| c || c | c |}
           \hline
     Ranking & Candidate & Score \\
       \hline
1 & Bernie Sanders & 8\\
2 & Gary Johnson & 7\\
3 & Jill Stein & 6\\
4 & Marco Rubio & 5\\
5 & John Kasich & 4\\
6 & Ted Cruz & 3\\
7 & Hillary Clinton & 2\\
8 & Donald Trump & 1\\\hline
    \end{tabular}}
    \caption{Coombs' Method}
       \label{table:coombs}
       \end{table}

\subsection{Copeland's Rule Results}
Copeland's Rule is the only mechanism which will elect a Condorcet winner, which is why Bernie Sanders also won this mechanism (Table \ref{table:copeland}).
\\
\indent On the other end, Donald Trump loses every pairwise match up, giving him a score of -7. If our data and computations are a representation of the population, this is making a very strong statement about the current state of the United States presidential election system, since it elected a candidate that should lose every direct match up with another top candidate.

    \begin{table}[h!]
    \centering
    {\begin{tabular}{| c || c | c |}
           \hline
     Ranking & Candidate & Score \\
       \hline
1 & Bernie Sanders & 7\\
2 & Gary Johnson & 5\\
3 & Hillary Clinton & 3\\
4 & Jill Stein & 1\\
5 & Marco Rubio & -1\\
6 & Ted Cruz & -3\\
7 & John Kasich & -5\\
8 & Donald Trump & -7\\\hline
    \end{tabular}}
    \caption{Copeland's Rule}
       \label{table:copeland}
       \end{table}

\subsection{Borda Count Results}
While Bernie Sanders continues to win in the Borda Count mechanism, Gary Johnson places second, which occurs in a total of four mechanisms (tables \ref{table:coombs}, \ref{table:copeland}, \ref{table:borda} \ref{table:moderation}). Since Gary Johnson only received approximately one percent of the popular vote in the 2016 presidential election \cite{johnson}, the outcomes of our mechanisms indicate that Gary Johnson was a fairly strong candidate.

    \begin{table}[h!]
    \centering
    {\begin{tabular}{| c || c | c |}
           \hline
     Ranking & Candidate & Score \\
       \hline
1 & Bernie Sanders & 4108\\
2 & Gary Johnson & 3929\\
3 & Hillary Clinton & 3556\\
4 & Marco Rubio & 3520\\
5 & Donald Trump & 3326\\
6 & Jill Stein & 3260\\
7 & John Kasich & 3102\\
8 & Ted Cruz & 3018\\\hline
    \end{tabular}}
    \caption{Borda Count}
       \label{table:borda}
       \end{table}

\subsection{Moderation Rule Results}
Since this mechanism is a hybrid of Coombs' Method and Ranked Choice, it shared the same winner once again in Bernie Sanders (Table \ref{table:moderation}). One of the nice features that our Moderation Rule has, is that it guarantees that we will not see a winner in candidates which perform very poorly in either Coombs' Method or Ranked Choice. Notice that Donald Trump places last in Coombs' Method Approval Voting{table:coombs}), but places second in Ranked Choice (Table \ref{table:rc}). Conversely, Jill Stein places last in Coombs' Method, but third in Ranked Choice. The Moderation Rule incorporates indicators from both mechanisms, resulting in Donald Trump and Jill Stein as the bottom two ranks in the outcome.

    \begin{table}[h!]
    \centering
    {\begin{tabular}{| c || c | c |}
           \hline
     Ranking & Candidate & Score \\
       \hline
1 & Bernie Sanders & 8\\
2 & Gary Johnson & 7\\
3 & Ted Cruz & 6\\
4 & John Kasich & 5\\
5 & Hillary Clinton & 4\\
6 & Marco Rubio & 3\\
7 & Donald Trump & 2\\
8 & Jill Stein & 1\\\hline
    \end{tabular}}
    \caption{Moderation Rule}
       \label{table:moderation}
       \end{table}

\subsection{Candidate Polarization}
When observing the distributions of rankings per candidate in table \ref{figure:voter_rankings}, we notice a pattern between certain candidates' distributions. We can define a general notion of the polarity of a candidate by observing the extremes of ranks: a polar candidate should be both favorite and most hated by voters. The ranking distribution of candidates Donald Trump and Hillary Clinton demonstrate this concept -- they have many ranks of 1 and 8 and very few middle ranks of 4 and 5. On the other hand, the distributions of candidates like Marco Rubio and Gary Johnson represent the opposite effect -- they have many votes which rank them between 3 and 6, and very few ranks of 1 and 8.
\\
\indent We used these concepts as motivation to develop a new metric which takes into account the polarity of an unacceptable candidate, which had already been outlined in algorithm 2. This mechanism penalized candidates with polar voting rankings such as Donald Trump, while rewarded candidates like Marco Rubio and Gary Johnson as seen in table \ref{table:polar}. To provide an intuition behind the function of the Polarization Rule, we claim that Gary Johnson is a clear winner because he frequently occurs on voters' acceptable lists (Table \ref{table:av}). Also, for votes where he is not on the acceptable list, Gary Johnson seldom occurs toward the bottom of the rankings, granting him a low penalty. The outcome here is that by electing Gary Johnson, the negative reaction by voters would be minimized.

    \begin{table}[h!]
    \centering
    {\begin{tabular}{| c || c | c |}
           \hline
     Ranking & Candidate & Score \\
       \hline
1 & Gary Johnson & -1656\\
2 & Marco Rubio & -2323\\
3 & Bernie Sanders & -2500\\
4 & Jill Stein & -2761\\
5 & John Kasich & -3779\\
6 & Ted Cruz & -4546\\
7 & Hillary Clinton & -4799\\
8 & Donald Trump & -6058\\\hline
    \end{tabular}}
    \caption{Polarization Rule}
       \label{table:polar}
       \end{table}

\section{Discussion}
\subsection{Truthfulness}
Whenever voting mechanisms are applied, the issue of truthfulness must be discussed. For the purposes of this project, we assumed that the ranked list participants submitted was based on their truthful preferences. Participants did not know the different voting mechanisms that would be used in this project, so it is unlikely that they would have submitted an untruthful ranked list. However, some participants in our survey commented that they did not know some of the candidates that we had them rank. This could affect how the ranked lists were made because ignorance of a candidate could theoretically cause them to appear more towards the middle of people's lists. However, ignorance about some of the candidates could also be representative of the normal United States election, where voters typically do not know everything about all of the candidates.\\
\indent
One interesting finding regarding truthfulness appeared in the question asking for a first round vote in the plurality with runoff mechanism. Around 20 percent of our participants voted for a different candidate than the first choice in their ranked list. Additionally, the most compelling switch happened when a voter had a first choice of Donald Trump but submitted a vote for Bernie Sanders. Our hypothesis for this is based on further analysis of these voters that switched, because many of these voters who ranked Donald Trump first also approved of Bernie Sanders. From this, we theorize that since Trump, Sanders, and Clinton were the three leading candidates in the 2016 Presidential Election, these voters wanted to ensure that the two candidates who ended up in the runoff were both candidates that they found acceptable. Therefore, if they assumed Trump would make it into the final runoff without their vote, they would vote for another popular candidate they approved of to make sure the resulting winner was acceptable to them. This is just conjecture, however, and without gathering more data we cannot conclude anything.

\subsection{Selecting An Optimal Mechanism}
What factors are most important when deciding on an election mechanism for a democracy? Based on our research and the surrounding literature review, we arrive at a few themes. \newline\newline
\textit{Comprehensibility}  

While there are multiple reasons why plurality is the most widely accepted voting mechanism, the most obvious reason is its simplicity. A properly functioning democracy must have a comprehensible voting mechanism and may potentially need to sacrifice effective fair voting in order to be simple enough for all to understand. Ideally, democracies should not want to make an election mechanism so complicated that only an elite few can understand or critique on the election. Failing to do this can potentially frustrate voters or raise suspicion in the reliability of their voting system. Thus, most alternatives to plurality vote involve simple, intelligible modifications. These include the Borda Count, the primary system, approval voting, or plurality rule with runoff.\cite{levin}\cite{arrow}  \newline\newline
\textit{Ease Of Implementation} 

Similar to having a comprehensible voting mechanism, an easily implementable voting system is critical. In a country like America, voting occurs in places of various technological literacy and sophistication. Additionally, citizens vote through various methods like absentee ballots, mail-in voting, and in-person voting, depending on their job or background.

A proper voting mechanism should not be too complicated to be correctly implemented. Additionally, it must not threaten voter turnout or increase voter disenfranchisement because of a difficulty in participation. Finally, a realistic voting mechanism should be auditable at scale, both to maintain confidence in the system and also to protect against sabotage. Regarding our Moderation Rule and Polarization Rules, it is easy to understand what these rules are doing, but much harder for the everyday voter or voting official to debate the pros and cons of the strategy. \\ \\
\textit{Truthfulness} \\
\indent
Similar to Section 7.1, we ideally want voters to vote in a truthful fashion, rather than incentivizing citizens to change their preferences to adjust to the complications of the system. As we've addressed earlier, it is very difficult to ascertain truthfulness with many election mechanisms, including ours. This would be an excellent subject for further study. \\ \\
\textit{Reducing Partiality}\\
\indent
Another important consideration is making sure that candidates are not incentivized to ignore certain segments of the voting population to achieve victory. Thus, for example, in America, an election mechanism should disincentivize politicians to campaign solely in cities or conversely campaign solely in rural areas.  \\ \\
\textit{Reducing Polarization}\\
\indent
Finally, as we cover in this paper, we believe that building election mechanisms that reduce the election of polarized candidates can increase stability and improve democracy. While there are certainly benefits to allowing for radical ideas and spirit in our government, the erosion of Chile's democracy \cite{chile}, as well as examples of other non-majority plurality rule countries, demonstrate the threat to stability certain election mechanisms can have. Stability essentially. In this way, our current two-party system usually disincentivizes intense polarization \cite{hall}; however, a few of our proposed methods might achieve this effect better. 

\section{Future Work}
Based on our findings in this project, our next steps would be to expand our mechanism implementations to other US elections as well as elections in other democracies. This would help to gather a better understanding of the average impact of our findings instead of limiting it to the scope of one election. Additionally, this would allow us to compare the impact of different mechanisms between different democracies. \\
\indent
Another direction would be to implement additional voting mechanisms such as quadratic voting. We wanted to keep our Qualtrics survey simple to maximize the number of responses. Therefore, this project did not include quadratic voting because it would have been more difficult for participants to quickly understand. If we were to send out an additional survey, it would be interesting to see the result of quadratic voting and compare it to the other mechanisms we have already implemented.\\
\indent
Lastly, many of the research papers in our literature review came from current professors at Stanford's Department of Political Science. If we were to continue working on this project, we would like to contact Andrew Hall, Shanto Iyengar, and Bruce Cain because their work deals with incentives in election mechanisms. Specifically, Andrew Hall's work \cite{hall}\cite{hallextremeprimaries} has many parallels with this project, which opens up the possibility for future collaboration.

\section{Conclusion}
In this paper, we outline various election mechanisms and evaluate them on the 2016 United States Presidential Election. To do this, we collected data on voter preferences using Mechanical Turk and extrapolated these findings to represent the population that voted in 2016. After analyzing these results, we also define two novel mechanisms with the purpose of reducing polarization. Finally, we identify key flaws in certain mechanisms and discuss the trade-offs between all of the implemented voting rules. The result of this paper demonstrates the impact a voting mechanism has on democratic elections. This analysis warrants a deeper discussion of the current implementation of the United States' electoral system. 


\begin{thebibliography}{99}

\bibitem{chile}
Valenzuela, J. Samuel, and Timothy R. Scully. "Electoral choices and the party system in Chile: Continuities and changes at the recovery of democracy." (1997): 511-527.

\bibitem{Norris}
Norris, Pippa. “Choosing Electoral Systems: Proportional, Majoritarian and Mixed Systems.” International Political Science Review, vol. 18, no. 3, July 1997, pp. 297–312, doi:10.1177/019251297018003005.

\bibitem{california}
J. Sides, “Can California’s New Primary Reduce Polarization? Maybe Not.,” The Monkey Cage, 27-Mar-2013. [Online]. Available: http://themonkeycage.org/2013/03/can-californias-new-primary-reduce-polarization-maybe-not/.

\bibitem{arrow}
K. Arrow, “Social Choices and Individual Values.” Cowles Commission for Research in Economics. 1951.

\bibitem{black}
Black, Duncan, The Theory of Committees and Elections. London: Cambridge University Press, 1953.

\bibitem{condorcet}
Condorcet, Marquis de, "Essay on the Application of Mathematics to the Theory of Decision Making," 1785. In Baker, K., ed., Condorcet: Selected Writings. Indianapolis: Bobbs-Merrill, 1976.

\bibitem{coombs}
Coombs, Clyde, A Theory of Data. New York: Wiley, 1964.

\bibitem{copeland}
Copeland, A. H., "A 'Reasonable' Social Welfare Function," mimeo, University of Michigan Seminar on Applications of Mathematics to the Social Sciences, 1951

\bibitem{minviolations}
Ali, I., W. D. Cook, and M. Kress, "On theMinimum Violations Ranking of a Tournament," Management Science, June 1986, 32, 660–72.

\bibitem{myerson}
Myerson, Roger, "Incentives to Cultivate Favored Minorities under Alternative Electoral Systems," American Political Science Review, December 1993, 87, 856–69.

\bibitem{sen}
A. Sen, “The Possibility of Social Choice,” American Economic Review, vol. 89, no. 3, pp. 349–378, Jun. 1999.

\bibitem{hall}
A. B. Hall and D. M. Thompson, “Who Punishes Extremist Nominees? Candidate Ideology and Turning Out the Base in US Elections,” American Political Science Review, vol. 112, no. 03, pp. 509–524, Aug. 2018.

\bibitem{hallextremeprimaries}
A. B. Hall, “What Happens When Extremists Win Primaries?,” American Political Science Review, vol. 109, no. 01, pp. 18–42, Feb. 2015.

\bibitem{iyengar} 
S. Iyengar and M. Krupenkin, “The Strengthening of Partisan Affect: Strengthening of Partisan Affect,” Political Psychology, vol. 39, pp. 201–218, Feb. 2018.

\bibitem{pacuit}
Pacuit, Eric, "Voting Methods", The Stanford Encyclopedia of Philosophy (Fall 2017 Edition), Edward N. Zalta (ed.).

\bibitem{levin}
Levin, J., \& Nalebuff, B. (1995). An Introduction to Vote-Counting Schemes. Journal of Economic Perspectives, 9(1), 3-26. doi:10.1257/jep.9.1.3

\bibitem{rickard}
Rickard, S. J. (2017). Electoral Systems and Policy Outcomes. Oxford Research Encyclopedia of Politics. doi:10.1093/acrefore/9780190228637.013.267

\bibitem{lecture3}
Roughgarden, T. (2016, October 3). Strategic Voting. Retrieved December 6, 2018, from http://theory.stanford.edu/~tim/f16/l/l3.pdf

\bibitem{johnson}
“Gary Johnson 2016 presidential campaign,” Wikipedia.

\bibitem{exitpoll}
Exit Polls 2016. (2016, November 23). Retrieved December 6, 2018, from https://www.cnn.com/election/2016/results/exit-polls

\bibitem{cmulecture}
Procaccia, A. D. (2008, June 18). Lecture 6 Notes: Social Choice Theory [CMU Mathematical Foundations of AI]. Carnegie Mellon University.

\bibitem{homework}
Mei, D., \& Chuchro, R. (2018, December 5). CS 269I Exercise Set 9. CA, Stanford.

\end{thebibliography}

\clearpage

\onecolumn
\noindent
\LARGE \textbf{Appendix}

\begin{table*}[h!]
\centering
\resizebox{\textwidth}{!}{%
\begin{tabular}{|l|l|l|l|l|l|l|l|l|l|}
\hline
\textbf{Candidates/Mechanisms} & \textbf{Plurality} & \textbf{w/ Runoff} & \textbf{Approval} & \textbf{RC} & \textbf{Coombs} & \textbf{Copeland} & \textbf{Borda} & \textbf{Moderation} & \textbf{Polarization} \\ \hline
\textbf{Gary Johnson}          & 4                  & 6                  & 4                 & 4           & 2               & 2                 & 2              & 2                   & 1                     \\ \hline
\textbf{Hillary Clinton}       & 3                  & 3                  & 2                 & 3           & 7               & 3                 & 3              & 5                   & 7                     \\ \hline
\textbf{Jill Stein}            & 8                  & 8                  & 8                 & 8           & 3               & 4                 & 6              & 8                   & 4                     \\ \hline
\textbf{Donald Trump}          & 1                  & 2                  & 3                 & 2           & 8               & 8                 & 5              & 7                   & 8                     \\ \hline
\textbf{Marco Rubio}           & 6                  & 5                  & 5                 & 6           & 4               & 5                 & 4              & 6                   & 2                     \\ \hline
\textbf{Bernie Sanders}        & 2                  & 1                  & 1                 & 1           & 1               & 1                 & 1              & 1                   & 3                     \\ \hline
\textbf{Ted Cruz}              & 7                  & 7                  & 6                 & 7           & 6               & 6                 & 8              & 3                   & 6                     \\ \hline
\textbf{John Kasich}           & 5                  & 4                  & 7                 & 5           & 5               & 7                 & 7              & 4                   & 5                     \\ \hline
\end{tabular}%
}
\caption{Table of mechanism rankings}
\label{table:all_rankings}
\end{table*}


\end{document}